# A NODE DEPLOYMENT MODEL WITH VARIABLE TRANSMISSION DISTANCE FOR WIRELESS SENSOR NETWORKS

Fan Tiegang[1] and Chen Junmin[2]

[1]Key Laboratory of Machine Learning and Computational Intelligence, College of Mathematics and Information Science, Hebei University, Baoding 071002, Hebei, China
[2]College of Mathematics and Information Science ,Hebei University, Baoding, 071002, Hebei, China

***ABSTRACT***

*The deployment of network nodes is essential to ensure the wireless sensor network's regular operation and affects the multiple network performance metrics, such as connectivity, coverage, lifetime, and cost. This paper focuses on the problem of minimizing network costs while meeting network requirements, and proposes a corona-based deployment method by using the variable transmission distance sensor. Based on the analysis of node energy consumption and network cost, an optimization model to minimize Cost Per Unit Area is given. The transmission distances and initial energy of the sensors are obtained by solving the model. The optimization model is improved to ensure the energy consumption balance of nodes in the same corona. Based on these parameters, the process of network node deployment is given. Deploying the network through this method will greatly reduce network costs.*

## 1. INTRODUCTION

With the development of wireless communication and microelectronics technology, wireless sensor networks have received more attention from researchers and many industries. It is used in many practical applications, including battlefield surveillance, forest fire monitoring, crop detection, environmental monitoring, medical insurance, industrial automation, etc. It changes the way people sense the world to a large extent, and at the same time, it also changes humanity's lifestyle. The location of the sensor and the network's operation mode are related to the regular operation and performance of the network. Network deployment is one of the essential tasks for wireless sensor networks.

Many node deployment methods have been proposed. They can be divided into static node deployment method and movable node deployment method based on whether the node is movable. The static deployment method includes deterministic deployment and random deployment. The movable node deployment method includes centralized deployment and distributed deployment. Different deployment methods focus on different network performance, mainly network coverage, connectivity, time delay, network lifetime, energy efficiency, network cost, etc. There is a close correlation between these metrics. Since sensors are generally powered by batteries and the energy is limited, saving energy and extending network lifetime has become an important research content for network node deployment.

With the continuous expansion of WSNs detection area, more and more attention is paid to the network cost. How to deploy the network to minimize the network cost while meeting performance requirements has become an important research content. There are not many works on network cost, and the network cost concerned mainly refers to the number of nodes. In addition to the number of sensors, many factors affect the network cost, such as the area of the detection area, energy efficiency, and network lifetime. As far as I know, there is no research to discuss network cost based on the above factors. Our goal is to design a network structure and network node deployment scheme that minimizes the cost of the network while meeting the given performance requirements.

This paper gives an optimization model to minimize the network kcost, and proposes a deployment method. To sum up, the main contributions are presented as follows.

(1). A "corona + cluster" network structure is proposed. The width and cluster size of each corona is different.

(2). By analyzing the energy consumption and network cost, an optimization model is proposed to minimizing cost per unit area.

(3). The deployment parameters are obtained by solving and improving the model, and the deployment steps of the multi-sink network are given.

## 2. RELATED WORK

Effective deployment of Sensor Nodes(SNs) is a significant point of concern as the performance and lifetime of any WSN primarily depend on it. Various models have been proposed by researchers to deploy SNs in large-scale open regions. Network coverage and connectivity are the fundamental concerns of node deployment. The paper [1] describes relevant deployment strategies to meet coverage requirements. Network connectivity and coverage are closely related, which mainly depends on the relationship between sensing distance and communication distance. Paper [2][3] give relevant conclusions. In the large-scale field or the inaccessibility/harshness field, random deployment becomes only one option. There are two classes random deployment strategies, simple strategies, and compound strategies [4]. One of the main research contents of random deployment is determining the distribution function of nodes to ensure network connectivity and coverage. The paper[5] gave an analytical expression for estimating the average minimum number of nodes for getting full connection of networks. The paper [6] proposed a probabilistic approach to compute the covered area fraction at critical percolation for both SCPT and NCPT problems. In [7], a novel framework was proposed for solving optimal deployment problems for randomly deployed and clustered WSNs. The percolation theory is adopted to analyze the degree of connectivity when the targeted degree of partial coverage is achieved.

Many researcher focus on the relocation of mobile sensors after random deployment. The paper [8] proposed the Virtual Force Algorithm (VFA) for clustering-based networks. In this method, the cluster head node calculates the coverage of the area and sends instructions to the cluster members to move to the corresponding position accordingly. Paper [9] proposed a deployment algorithm for nodes with limited mobility. Limited mobility means that the maximum distance a node can move is limited. The paper [10] proposed a non-uniform deployment method that can approximate energy balance consumption. It is assumed that all nodes are deployed in a circular area, the sink is located at the center of the circle, the transmission distance of the sensing node is the same, and the node's energy consumption includes data reception and transmission. Paper [11] proposed a non-uniform deployment method based on unequal clusters to optimize the Energy-Consuming Rate. This method combines the advantages of unequal clusters and non-uniform deployment and adopts a node sleep mechanism. It not only improves energy efficiency but also balances node energy consumption and avoids the energy holes problem. The paper [12] studied the problem of unbalanced energy consumption in a uniformly distributed network and proposed two random relay node deployment methods. The first method aims to balance the energy consumption rate of relay nodes, thereby extending the network's lifetime. This method does not guarantee the connection with the sensor node. The second method considers lifetime and connectivity. The article discusses both single-hop and multi-hop scenarios.

The paper [13] proposed a novel modeling solution capable of representing a wide variety of scenarios, from totally random to planned stochastic node deployment, in heterogeneous sensor networks. Using only about 2% of H-sensors and deploying nodes by using the P- and Q-models to distribute the L-sensors around the H-sensors deployed with a repulsive model, they observed essential characteristics of the network topology, such as low average path length and high clustering coefficient.The paper [14] gave a survey about clustering and cluster-based multi-hop routing protocols. Some parameters were given to evaluate the properties of the different methods. The studied methods are classified into four categories: classical approaches, fuzzy-based approaches, metaheuristic-based approaches, and hybrid metaheuristic- and fuzzy-based approaches. In each category of the classification, criteria and parameters are presented according to the type of methodology. The authors [15] gave a efficient coding techniques algorithm for cluster-heads communication in wireless sensor networks to improve the Bit-error rate. The paper [16] proposed a new cluster-based algorithm to minimize energy consumption in WSNs.

In the wireless sensor network, the uneven energy consumption of the nodes significantly shortens the network lifetime and causes waste. An essential idea to solve this problem is to equip nodes with different energy. Paper [17] studied the benefits of multi-level batteries for prolonging the network's lifetime, showing that under the total energy budget, the network lifetime under the multi-level battery allocation method is much longer than that under the same battery allocation. Paper [18] further studied the battery distribution method. The problem can be described as: There are m types of batteries, and the power and cost are known. Different types and numbers of batteries can be packaged and distributed to the node. Assuming that there M battery packs, determine the number of each type battery in each pack to maximize the network lifetime. In order to balance the energy consumption of sensors, the mixed/hybrid transmission schemes appeared in many papers [19]. Each sensor can adjust its transmission power level and alternates between direct transmission mode (sending data directly to sink without using any relay node) and hop-by-hop transmission mode (forwarding data to next-hop neighbors).

According to the energy consumption model of node data transmission, energy consumption can be changed by adjusting the transmission distance. Some deployment methods have been developed to balance the energy consumption of nodes by adjusting the nodes' data transmission distance. The paper [20] discussed the problem of balancing the energy consumption of nodes by changing the transmission distance, and provided an iteration-based method for determining the transmission distance. This paper proposed an improved corona model with levels to investigate the transmission range assignment strategy used to maximize the lifetime of wireless sensor networks. The author proved the problem of searching optimal transmission range lists is a multi-objective optimization problem, and that is also NP hard. The author [21] proposed a mixed integer programming formulation, i.e., SPSRC, that combines all design issues in a single model. The SPSRC finds the sensors' and sinks' optimal locations, active/standby periods of the sensors, and the data transmission routes from each active sensor to its assigned sink.

In this paper, we introduce a node deployment strategy based on variable transmission distance. Based on the network structure of "corona + cluster," an optimization model to minimize network cost is proposed.

## 3. NETWORK STRUCTURE

It is assumed that there is a circular detected area with a radius of R. The base station is located at the center of the area. The sensor has a fixed sensing radius and adjustable data transmission distance. The problem to be solved is how to deploy sensors to minimize network cost while meeting relevant design requirements.

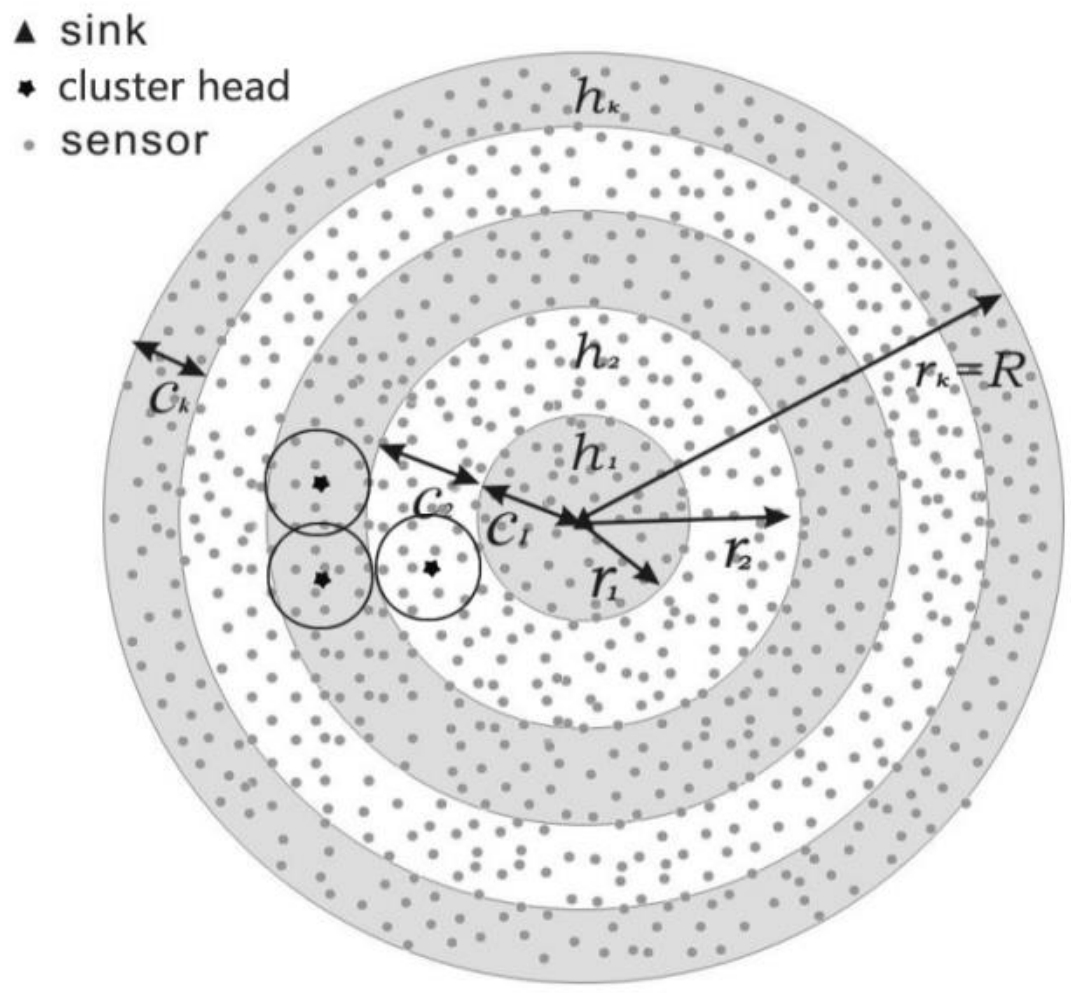

Figure 1. The network structure of " corona + cluster"

The sensors can be placed uniformly by random throwing, and the density of the nodes can be obtained on the sensing radius. The monitoring area is divided into several coronas by concentric circles. The transmission distance of the sensor in the corona is equal to the corona's width where it is located. As shown in Figure 1, $h_i$ represents the ith corona, and the corresponding circle radius is represented by $r_i$. The width of the ith corona $c_i = r_i - r_{i-1}$. The sensors in the same corona form multiple clusters. The number of clusters is related to the radius of the corresponding circle. The cluster head nodes are generated from the nodes in the cluster and are continuously rotated. Sensors periodically sense the environment, generate data, and send the data to cluster head nodes in the same cluster through a single hop. The cluster head nodes receive the cluster node's data, integrate the cluster data, receive the data sent by the cluster head node of the previous corona, and transmit the data to the cluster head node of the next corona.

Since the transmission distances of sensor nodes in different coronas are different, we use an energy consumption model and introduce the following symbols:

$e_0$: The energy consumption of generating unit data,

$e_1$: The energy consumption of electrical devices receiving or sending unit data,

$e_2$: Amplification factor when sending data

$e_3$: The energy consumption of integrated unit data

The energy consumption formula for sending unit data is: $e_1+e_2d^2$ , where $d$ is the sending distance. The sensor's energy consumption per unit time is:

$$E = e_0x + e_1y + (e_1 + e_2d^2)z + e_3n,$$

where $x, y, z, n$ indicate the corresponding amount of data, respectively.

Since the energy consumption of cluster management and cluster head generation is far less than that of data transmission, it can be ignored. Of course, it can also be set as a fixed value independent of data volume, namely the fixed energy consumption per unit time, which has almost no impact on our optimization model's solution.

The cost model of the sensor is $a+bE$, where $a$ is the hardware cost of sensor, $b$ is the unit energy cost, $E$ is the energy value fixed at the node. It is known that the energy consumption of each corona node is different. To effectively use the energy, we will adopt the method that assembles different initial energy for the nodes in different coronas. In this way, the costs of nodes in different coronas are different. The base station is powered by dedicated lines or solar energy, and its cost is fixed.

Network lifetime refers to the minimum lifetime of a node $T = \min\{T_i\}$, where $T_i$ is the lifetime of node i. In an ideal network, each node has the same lifetime. One of our design goals is that all nodes have approximately the same lifetime. As a design index, network lifetime should meet the application requirement. Network design lifetime is an important factor affecting network cost.

## 4. OPTIMIZATION MODEL

### 4.1. Analysis and Model

Firstly, the sensor's energy consumption is analyzed. We divide the energy consumption into two parts: (1) Intra-cluster energy consumption, including energy consumption of sensor, sensing the environment and transmitting data to the cluster head node, and energy consumption of the cluster head node that receives, integrates the data in the cluster and sends to next cluster head node. (2)Inter-cluster energy consumption refers to the energy consumption when the cluster-head node receives data from the previous cluster-head node and sends it to the next cluster head node.

The number of nodes in the ith corona is $N_i$, then

$$N_i = \rho\pi(r_i^2 - r_{i-1}^2)$$

Where, $\rho$ is a node density that ensures to meet the coverage requirements of the monitoring area. It has been proved in literature [22] that at least $2\pi r_i / c_i$ clusters are needed to cover the ith corona. We can calculate the proportion of cluster-head nodes to all nodes in the ith corona.

$$p_i = \frac{2\pi r_i / c_i}{\rho\pi(r_i^2 - r_{i-1}^2)} = \frac{2r_i}{\rho c_i^2(2r_i - c_i)},$$

The energy consumption per unit time of each cluster-head node in ith corona is:

$$E_{ch_i} = e_0 l + (\frac{1}{p_i} - 1)e_1 l + e_3 l\frac{1}{p_i} + ml(\mathrm{e}_2\, c_i^{\,2} + e_1)\frac{1}{p_i},$$

where l is the amount of data generated by a sensor per unit time. In the above formula, the first term is the energy consumption for generating data, the second is the energy consumption of receiving data from other nodes in the cluster, the third is the energy consumption of integrating data, and the fourth is the consumption for sending the integrated data to the next cluster head node, where m is the data compression factor. In many non-uniform clustering algorithms, clusters far from the base station are larger than clusters close to the base station [11]. To ensure sending data to the cluster head node of the next corona, transmission distance of the cluster head node is set as the corona width.

The energy consumption per unit time of sensor in ith corona is:

$$E_{nch_i} = e_0 l + l(\mathrm{e}_2\, c_i^{\,2} + e_1),$$

The first term is the energy consumption for generating sensory data. The second is the energy consumption of sending the data to the cluster head node. Based on the above two formulas, the energy consumption of all nodes in ith corona per unit time can be obtained.

$$\begin{aligned} E_{\mathrm{intra}_i} &= N_i p_i E_{ch_i} + N_i(1-p_i)E_{nch_i} \\ &= N_i p_i e_0 l + N_i(1-p_i)e_1 l + N_i e_3 l + N_i ml(\mathrm{e}_2\, c_i^{\,2} + e_1) + N_i(1-p_i)e_0 l + N_i(1-p_i)l(\mathrm{e}_2\, c_i^{\,2} + e_1) \\ &= N_i l\{e_0 + [2(1-p_i)+m]e_1 + (1-p_i+m)\,\mathrm{e}_2\, c_i^{\,2} + e_3\} \end{aligned}$$

The inter-cluster energy consumption per unit time of all cluster head nodes in ith corona is:

$$\begin{aligned} E_{\mathrm{inter}_i} &= \rho\pi(R^2 - r_i^2)mle_1 + \rho\pi(R^2 - r_i^2)ml(e_1 + e_2 c_i^{\,2}) \\ &= \rho\pi(R^2 - r_i^2)ml(2e_1 + e_2 c_i^{\,2}) \end{aligned}$$

In particular, since the cluster head nodes in the outermost corona does not receive data from other coronas, $E_{\mathrm{inter}_k} = 0$. The total energy consumption of all nodes in ith corona per unit time is:

$$E_i = E_{\mathrm{intra}_i} + E_{\mathrm{inter}_i},$$

It can be seen that the total energy consumption is different under different corona numbers and corona widths.The average energy consumption per unit time of each node in ith corona is:

$$\bar{E}_i = \frac{E_i}{N_i},$$

Assuming that the network design lifetime is $T$, the average energy consumption of each node for cluster management (including cluster head generation, information exchange in the cluster, etc.) per unit time is a fixed value $E^*$. The energy assembled by each node in the corona should be:

$$E_{p_i} = (\bar{E}_i + E^*)T,$$

According to the previous cost model, the total network cost can be obtained as:

$$C = a\pi R^2 \rho + b\sum_{i=1}^{k} N_i E_{p_i} + c^*,$$

The first term of the above equation is the hardware cost of all sensors, the second term is the cost of the total energy assembled on the sensors, and the third term $c^*$ is the cost of the base station. It can be seen that the total cost is different under the different number of coronas and corona widths. Therefore, it is hoped to find a suitable number of coronas and corona width to minimize the total cost.Further, theCost Per Unit Area(CPUA) can be obtained :

$$\begin{aligned} CPUA &= \frac{C}{S} \\ &= \frac{a\pi R^2 \rho + b\sum_{i=1}^{k} N_i E_{p_i} + c^*}{\pi R^2} \\ &= a\rho + b\rho E^* T + Tb\sum_{i=1}^{k} E_i \frac{1}{\pi R^2} + \frac{c^*}{\pi R^2} \end{aligned}$$

The meaning of the above formula is also obvious. With the CPUA as the target function, the following optimization model is obtained:

$$\begin{aligned} \min \quad & CPUA \\ s.t. \quad & d_1 \le c_i \le d_2 \\ & c_i = r_i - r_{i-1}, i = 2 \cdots k \\ & c_1 = r_1 \\ & r_k = R \end{aligned}$$

where, $d_1$ and $d_2$ is the lower and upper limits of node transmission distance. When the detection region is given, namely R is the determined value, the objective function of the above model can be replaced by the total network cost, and the optimal solution of the decision variable is the same.

When the relevant parameter values( $\rho$, $R$, $T$, etc.) of the model are given, minimizing the unit area cost CPUA (or total cost C) is equivalent to minimizing total energy consumption (the sum of energy consumption except for cluster management). The above model can be simplified as:

$$\begin{aligned} \min \quad & \sum_{i=1}^{k} E_i \\ s.t. \quad & d_1 \le c_i \le d_2 \\ & c_i = r_i - r_{i-1}, i = 2 \cdots k \\ & c_1 = r_1 \\ & r_k = R \end{aligned}$$

### 4.2. Solving the Model.

Given the relevant parameters, the above model can be solved. The values of the model parameters are as follows in Table 1.

Table 1 Model parameters

| **Parameters** | **Values** | **Parameters** | **Values** |
|---|---|---|---|
| $R$ | 200m | $e_0$ | 50nJ/bit |
| $\rho$ | 0.0318 | $e_1$ | 50nJ/bit |
| $l$ | 256bit/minute | $e_2$ | 10pJ/bit/m2 |
| $m$ | 0.1 | $e_3$ | 5nJ/bit |
| $d_1$ | 20m | $d_2$ | 80m |
| $a$ | 10$ | $b$ | 2$ |
| $T$ | 100000minutes | $c^*$ | 200$ |
| $E^*$ | 100nJ/minute | | |

Solve the model according to the following steps to determine the optimal number of coronas and the width of each corona.

(1) Determine the upper and lower limits of the number of coronas according to the upper and lower limits of the data transmission distance of the sensing node. Since , , , it can be known that the number of coronas is between 3-10.

(2) Solve the model under each corona number respectively to get the optimal corona width and minimum total energy consumption( or unit area cost) of each corona under the corresponding number of coronas.

(3) Compare the total energy consumption(or unit area cost) between different corona numbers, and take the corona number and corona width corresponding to the minimum value to be the final result.

The solution of the model are shown in Table 9. It can be seen that when the number of coronas is 6, the CPUA or total energy consumption is the lowest. The widths of the first corona to the sixth corona are: 58.5m, 42.3m, 33.9m, 25.3m, 20m, 20m, and the corona width decreases from the inside to the outside. This result is significantly different from the results of many different cluster algorithms[11]. In the EEUC algorithm, the cluster's size far away from the base station is larger than that close to the base station. The main reason is that our deployment strategy assembles different initial energies for the nodes.

Table 2 Energy consumption, cost and corona width under different number of coronas

| | k=3 | k=4 | k=5 | k=6 | k=7 | k=8 | k=9 | k=10 |
|---|---|---|---|---|---|---|---|---|
| Total energy consumption | 22817.15 | 21171.53 | 20598.77 | 20395.79 | 20517.97 | 20924.04 | 21573.38 | 22424.78 |
| CPUA | 0.6833738 | 0.657183 | 0.6480672 | 0.6448367 | 0.6467813 | 0.6532441 | 0.6635786 | 0.6771291 |
| h1 | 80 | 70.4 | 64.8 | 58.5 | 52.8 | 44.9 | 37.7 | 20 |
| h2 | 64.9 | 51.5 | 48 | 42.3 | 38 | 32.1 | 22.3 | 20 |
| h3 | 55.1 | 42.6 | 39 | 33.9 | 29.2 | 23 | 20 | 20 |
| h4 | | 35.5 | 27 | 25.3 | 20 | 20 | 20 | 20 |
| h5 | | | 20 | 20 | 20 | 20 | 20 | 20 |
| h6 | | | | 20 | 20 | 20 | 20 | 20 |
| h7 | | | | | 20 | 20 | 20 | 20 |
| h8 | | | | | | 20 | 20 | 20 |
| h9 | | | | | | | 20 | 20 |
| h10 | | | | | | | | 20 |

## 4.3. Model Improvement

From the results of the previous model, it can be seen that the width of the outer corona is smaller than the width of the inner corona. In the energy consumption analysis, we set the cluster head node's data transmission distance as the width of its corona, which will lead to uneven energy consumption for different corona's node. To solve this problem, we set the cluster head node's data transmission distance in ith corona as the width of (i-1)th corona $c_{i\text{-}1}$.

$$\tilde{E}_{ch_i}=\begin{cases} e_0 l+(\frac{1}{p_1}-1)e_1 l+e_3 l\frac{1}{p_1}+ml(\mathrm{e}_2\, c_1^{\ 2}+e_1)\frac{1}{p_1}, & i=1 \\ e_0 l+(\frac{1}{p_i}-1)e_1 l+e_3 l\frac{1}{p_i}+ml(\mathrm{e}_2\, c_{i\text{-}1}^{\ 2}+e_1)\frac{1}{p_i}, & i=2\cdots k \end{cases},$$

$$\tilde{E}_{\mathrm{int}er_i}=\begin{cases} \rho\pi(R^2-r_i^{\ 2})ml(2e_1+e_2c_1^{\ 2}), & i=1 \\ \rho\pi(R^2-r_i^{\ 2})ml(2e_1+e_2c_{i\text{-}1}^{\ 2}), & i=2\cdots k \end{cases}$$

By replacing $E_{ch_i}$ and $E_{\mathrm{int}er_i}$ in *CPUA* with the $\tilde{E}_{ch_i}$, and $\tilde{E}_{\mathrm{int}er_i}$, we get the $CPUA$, and the improved optimization model can be obtained:

$$\begin{aligned} \min \quad & CPUA \\ s.t. \quad & d_1 \le c_i \le d_2 \\ & c_i = r_i - r_{i-1},\ i = 2 \cdots k \\ & c_i \le c_{i-1},\ i = 2 \cdots k \\ & c_1 = r_1 \\ & r_k = R \end{aligned}$$

Constraint $c_i \le c_{i-1}$ is added to the model to ensure that the corona width changes regularly.

Resolve the model, and the CPUA and total energy consumption under different corona numbers are obtained . It can be seen that compared with the model solution before the improvement, the optimal value has increased corresponding to the same corona number. This is because of the increasing data transmission distance of cluster head nodes. By comparison, the cost is still the lowest when the number of coronas is 6. The corresponding corona width from the first corona to the sixth corona is 49.8m, 45.7m, 36.8m, 27.7m, 20m, 20m, respectively.

Table 3 Energy consumption and cost under different corona numbers (improved model)

| | k=3 | k=4 | k=5 | k=6 | k=7 | k=8 | k=9 | k=10 |
|---|---|---|---|---|---|---|---|---|
| Total energy consumption | 23062.72 | 21398.65 | 20798.46 | 20581.71 | 20677.48 | 21046.11 | 21634.69 | 22424.78 |
| CPUA | 0.6872822 | 0.6607977 | 0.6512453 | 0.6477956 | 0.6493199 | 0.6551868 | 0.6645543 | 0.6771290 |

## 5. NODE DEPLOYMENT BASIC STEPS

By solving the optimization model, we get the relevant parameters value for node deployment:

(1) The number of coronas $k$ and the width of each corona $c_i$;

(2) The number of sensors in each corona $N_i$;

(3) The proportion of cluster head nodes in each corona $p_i$; (4) The initial energy to be assembled for each node in each corona $E_{p_i}$.

The basic steps for node deployment are given below:

(1) According to the obtained corona width, the detected area is divided into coronas,

(2) Prepare s sensing nodes with initial energy f for each corona,

(3) Throw the corresponding nodes evenly on each corona,

(4) Each corona generates cluster heads, forms clusters, and starts data collection and transmission.

The fourth step above includes the generation and update mechanism of cluster heads and the data routing method between cluster heads, which has an important impact on the normal operation and energy consumption of the network. The previous optimization model does not cover these contents, which becomes our main work in the next step.

## 6. CONCLUSIONS

A node deployment method with variable transmission distance is proposed to minimize the network cost. We adopt the network structure based on "corona+cluster". The data transmission distance of the sensor can be adjusted. Sensors located in different coronas form clusters of different sizes. The nodes in different coronas are equipped with different initial energy. Based on the analysis of node energy consumption and network cost, an optimization model to minimize cost is proposed. The number of coronas and the optimal transmission distance of nodes can be determined by solving the model. It is found that the node's transmission distance gradually increases from the outer corona to the inner corona. For the energy consumption balance of the nodes in the same corona, the transmission distance of the cluster head node is revised, and the model is improved. On this basis, the steps of network node deployment are proposed

## 7. DISCUSSION

The sensors used have various characteristics in this study, such as adjustable data transmission distance and assembled with different initial energy. These characteristics appeared singly in the existing studies only, and the application of these characteristics can effectively reduce the network cost and improve the network performance. New features can be added in future work as the technology evolves. According to the conclusion of this paper, multi-sink network should be deployed for a large detected area. In this paper, the network lifetime is the ideal design lifetime, in general, while the actual network lifetime will be less than it. The generation and updating mechanism of clusters and the data routing algorithm have an important influence on network lifetime. These two elements will be focused on in future work.

## 8. ACKNOWLEDGMENTS

The paper sincerely thank the anonymous reviewers and the associate editor for valuable and constructive comments to improve the manuscript's quality and organization. This work is supported by grants from the Science and Technology Research Projects of Higher Schools(205020517502) in Hebei Province, China.

## AUTHORS

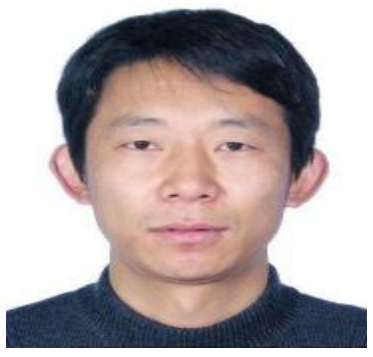

Fan Tiegang is an associate professor in the software engineering department of Hebei University. His research interests include optimization model, wireless sensor networks, and machine learning. He has published papers in journals and conferences in the area of wireless communications and artificial intelligence, and has been involved in projects with public funding.

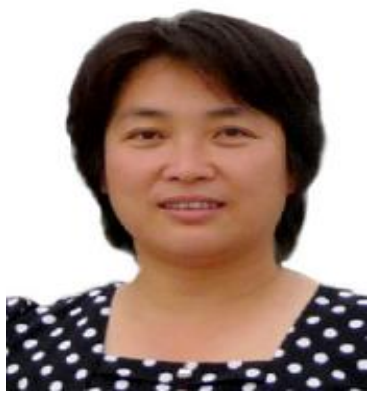

Chen Junmin is an associate professor in the department of mathematics of Hebei University. Her research interests include the optimization theory, the convergence of algorithms for variational inequalities. She has published papers in journals and conferences in the area of variational inequalities, and has been involved in one national project and two provincial projects.